\documentclass[12pt]{article}
\usepackage{graphicx} 
\usepackage[margin=2.7cm]{geometry}

\newcommand{\be}{\begin{equation}}
\newcommand{\ee}{\end{equation}}
\newcommand{\af}{{ \lambda }} 
\newcommand{\qhat}{ {\hat Q} } 
\newcommand{\lamroot}{{ \sqrt{\lambda} }} 
\newcommand{\yp}{{ y_k }} 

\newcommand{\period}{{ \Delta \tau }} 
\newcommand{\mfont}{ \mathcal } 
\newcommand{\square}{\bullet} 

\begin{document} 

\centerline{\bf HILL'S EQUATION WITH SMALL FLUCTUATIONS: } 
\centerline{\bf CYCLE TO CYCLE VARIATIONS AND STOCHASTIC PROCESSES}

\bigskip 
\bigskip 

\centerline{Fred C. Adams$^{1,2}$ and Anthony M. Bloch$^{1,3}$} 
\bigskip
\centerline{\it $^1$Michigan Center for Theoretical Physics, 
University of Michigan, Ann Arbor, MI 48109} 
\bigskip
\centerline{\it $^2$Astronomy Department, University of Michigan, Ann Arbor, MI 48109} 
\bigskip
\centerline{\it $^3$Department of Mathematics, University of Michigan, 
Ann Arbor, MI 48109} 

\begin{abstract} 

Hill's equations arise in a wide variety of physical problems, and are
specified by a natural frequency, a periodic forcing function, and a
forcing strength parameter. This classic problem is generalized here
in two ways: [A] to Random Hill's equations which allow the forcing
strength $q_k$, the oscillation frequency $\af_k$, and the period
$(\period)_k$ of the forcing function to vary from cycle to cycle, and
[B] to Stochastic Hill's equations which contain (at least) one
additional term that is a stochastic process $\xi$. This paper
considers both random and stochastic Hill's equations with small
parameter variations, so that $p_k=q_k-\langle{q_k}\rangle$,
$\ell_k=\af_k-\langle{\af_k}\rangle$, and $\xi$ are all 
${\cal O}(\epsilon)$, where $\epsilon\ll1$. We show that random Hill's
equations and stochastic Hill's equations have the same growth rates
when the parameter variations $p_k$ and $\ell_k$ obey certain
constraints given in terms of the moments of $\xi$. For random Hill's
equations, the growth rates for the solutions are given by the growth
rates of a matrix transformation, under matrix multiplication, where
the matrix elements vary from cycle to cycle.  Unlike classic Hill's
equations where the parameter space (the $\af$-$q$ plane) displays
bands of stable solutions interlaced with bands of unstable solutions,
random Hill's equations are generically unstable. We find analytic
approximations for the growth rates of the instability; for the regime
where Hill's equation is classically stable, and the parameter
variations are small, the growth rate $\gamma$ = ${\cal O}(p_k^2)$ =
${\cal O}(\epsilon^2)$. Using the relationship between the 
($\ell_k,p_k$) and the $\xi$, this result for $\gamma$ can be 
used to find growth rates for stochastic Hill's equations. 

\end{abstract} 

$\,$ 

\newpage
\baselineskip=18pt

\noindent 
{\bf I. INTRODUCTION} 
\medskip 

Hill's equation is a second order periodic differential equation that
arises in many physical problems [22]. In addition to its original
application for lunar orbits [15], Hill's equation describes celestial
dynamics [7,8,21], orbit instabilities in dark matter halos [5],
particle production at the end of the inflationary epoch in the early
universe [17,18,19], the motion of a jogger's ponytail [16], and many
other physical systems. The goal of this paper is to consider the
addition of random perturbations to Hill's equation and thereby
generalize our current understanding.  Toward this end, we adopt the
following definitions:

\medskip 
\noindent
{\bf Definition 1.1:} A Classical Hill's equation has the form 
\be 
{d^2 y \over dt^2} + [ \af + q \qhat (t) ] y = 0 \, , 
\label{classical} 
\ee 
where the parameters $(\af,q)$ are constant and where the barrier
shape function $\qhat(t)$ is periodic, so that $\qhat (t + \period) =
\qhat(t)$, where $\period$ is the period.  For the sake of
definiteness we take $\period = \pi$ and normalize the barrier shape
function so that $\int_0^{\pi}\qhat(t)dt=1$.

\medskip
\noindent
{\bf Definition 1.2:} A Hill's Equation is said to be Cycle-to-Cycle
Random (CC Random) if the differential equation is deterministic over
each interval of the periodic function $\qhat$, but the values of the
forcing strength $q_k$ and/or oscillation frequency $\af_k$ take on
different values for each periodic interval.  The CC Random Hill's
equation has the form 
\be 
{d^2 y \over dt^2} + [ \af_k + q_k \qhat (t) ] y = 0 \, , 
\label{ccrandom} 
\ee 
where the parameters ($\af_k, q_k$) vary from cycle to cycle, and are
drawn independently from well-defined distributions. The index $k$
labels the cycle.

In this work the period $\period$ is considered fixed; one can show
that cycle to cycle variations in $\period$ can be scaled out of the
problem and included in the variations of the $(\af_k,q_k)$ by
changing their distributions accordingly (see Theorem 1 of Ref. [2]). 

Periodic differential equations in this class can be described by a
discrete mapping of the coefficients of the principal solutions from
one cycle to the next.  The transformation matrix takes the form
\be
{\mfont M}_k = \left[ 
\matrix{h_k & (h_k^2 - 1)/g_k \cr g_k & h_k} \right] \, , 
\label{mapzero} 
\ee
where the subscript denotes the cycle. The matrix elements for the 
$kth$ cycle are given by  
\be
h_k = y_1 (\pi) \qquad {\rm and} \qquad g_k = {\dot y}_1 (\pi) \, , 
\label{hk}
\ee
where $y_1$ and $y_2$ are the principal solutions for that cycle. 
The index $k$ indicates that the quantities $(\af_k, q_k)$, and hence
the solutions $(h_k, g_k)$, vary from cycle to cycle. Throughout this
work, the random variables are taken to be independent and identically
distributed (iid).

Note that the matrix in equation $(\ref{mapzero})$ has only two
independent elements (not four). Since the Wronskian of the original
differential equation (\ref{ccrandom}) is unity, the determinant of
the matrix map (\ref{mapzero}) must also be unity, and this constraint
eliminates one independent element. In addition, this paper
specializes to the case where the periodic functions $\qhat(t)$ are
symmetric about the midpoint of the period.  This property implies
that $y_1(\pi)$ = ${\dot y}_2 (\pi)$ = $h_k$, which eliminates a
second independent matrix element [1,22]. These two constraints imply
that $y_2(\pi)$ = $(h_k^2-1)/g_k$, resulting in the form for the
matrix given by equation (\ref{mapzero}).

The growth rates for Hill's equation (\ref{ccrandom}) are determined by
the growth rates for matrix multiplication of the matrices 
${\mfont M}_k$ given by equation (\ref{mapzero}).  Here we denote the
product of $N$ such matrices as ${\mfont M}^{(N)}$, and the growth
rate $\gamma$ is defined by
\be
\gamma = \lim_{N \to \infty} {1 \over N} 
\log || {\mfont M}^{(N)} || \, . 
\label{growbasic}
\ee
This result is independent of the choice of the norm $|| \cdot ||$, 
and the limit exists almost surely, as shown in previous work [12,13,20].

\medskip
\noindent 
{\bf Definition 1.3:} A Stochastic Hill's Equation is a Hill's equation
with the inclusion of one or more additional terms that is a
stochastic process. In this paper, the stochastic Hill's equation is
written as a Langevin equation. For example, one can write stochastic 
Hill's equations in the form 
\be
{d^2 y \over dt^2} + [ \af + q \qhat (t) ] y = \xi \, , 
\label{stochastone} 
\ee
or 
\be
{d^2 y \over dt^2} + [ \af + \left( q + \xi \right) \qhat (t) ] y = 0 \, , 
\label{stochasttwo} 
\ee
where $\xi$ is a stochastic process and $(\af,q)$ are constant. 

Note that in the stochastic Hill's equation, we can evaluate the
stochastic term $\xi$ using the calculus of either Stratonovich or
Ito.  For purposes of this work, however, we only require that the
process $\xi$ has zero mean and finite variance.

The goal of this paper is to study both the CC Random Hill's equation 
and the Stochastic Hill's equation in the limit where the parameter 
variations and/or the stochastic terms are small. For the CC Random 
Hill's equation, the parameters are rewritten in the form 
\be
\af_k = \af + \ell_k \qquad {\rm and} \qquad 
q_k = q + p_k \, , 
\label{smallparameter} 
\ee
so that $(\af,q)$ are constant whereas the $\ell_k$ and $p_k$ are
small and are allowed to vary from cycle to cycle. The distributions
of the $\ell_k$ and the $p_k$ are assumed to 
have zero mean $\langle\ell_k\rangle$
= 0 = $\langle{p_k}\rangle$ and finite second moments 
$\langle \ell_k^2 \rangle$ and $\langle p_k^2 \rangle$. 

For both the CC Random and the stochastic Hill's equation, the
solution over a given cycle (period of the function $\qhat(t)$) can be
expanded such that 
\be
y (t) = y_0 (t) + \yp (t) \, ,
\label{smallfunction} 
\ee
where $y_0$ obeys the Classical Hill's equation (\ref{classical}), and
is the same for every cycle. The correction term $\yp(t)$ is, in
general, different for every cycle, due to the parameter variations
$(\ell_k,p_k)$ for the CC Random Hill's equation and due to different
realizations of the stochastic process $\xi$ for the Stochastic Hill's
equation (different samplings of the process). 

Next we introduce an ordering parameter $\epsilon \ll 1$ such that
\be
\yp = {\cal O}(\epsilon) , \qquad 
\ell_k = {\cal O}(\epsilon) , \qquad 
p_k = {\cal O}(\epsilon) , \qquad {\rm and} \qquad 
\xi = {\cal O}(\epsilon) \, .
\label{ordering} 
\ee
The results of this paper are correct to leading order in $\epsilon$. 

\medskip 
\noindent
{\bf Remark:} The subscripts `1' and `2' on the function $y(t)$ denote
the two principal solutions of any Hill's equation for a given cycle,
i.e., for given values of the parameters. The subscripts `0' and `$k$'
on $y(t)$ denote the leading order and first order parts (in $\epsilon$) 
of the function, respectively (see equation (\ref{smallfunction})).

To place this work in context, we note that a number of previous
studies have considered Mathieu's equation, a particular type of
Hill's equation, with stochastic forcing [25,26]. These papers show
that stochastic fluctuations can lead to instability in otherwise
stable systems. Other related work includes random operators in the
the Schr{\"o}dinger wave equation in quantum mechanics and population
variations in biomathematics [6,10,23], and stability studies of the
quasi-periodic Mathieu equation [28]. This body of existing work does
not, in general, provide closed-form expressions for the growth rates
of the differential equations.  Our previous work has studied the CC
Random Hill's equation in the regime where the forcing strengths are
large and the solutions are highly unstable [2], for the limit where
the periodic functions $\qhat(t)$ become (periodic) delta functions
[3], and for more general cases [4]; here we find closed-form
expressions for the growth rates of corresponding random matrices
(given by equation (\ref{mapzero})).  Examples of such analytic
results are rare in the literature -- the classic papers [9,12,13]
show the existence of growth rates, but present few examples (see also
Ref. [24]).

The main goal of this present work is to build a bridge between the
Stochastic Hill's equation and the CC Random Hill's equation.  This
paper focuses on the regime of small fluctuations as defined through
equation (\ref{ordering}).  We derive the conditions necessary for the
two types of Hill's equations to have the same growth rates (Theorem
2.3). We also find analytic expressions for the matrix elements
(Theorem 2.1) and the growth rates (Theorem 2.2) for CC Random Hill's
equations. The combination of these three results provides analytic
expressions for the growth rates for Stochastic Hill's equations.
Section III presents a special case of the problem where the natural
oscillation frequency $\af$ is constant and the forcing parameters
$q_k$ are small but fluctuating from cycle to cycle. The paper
concludes in Section IV with a summary and short discussion of the
results; representative applications are described in the appendices.

%\newpage 
\bigskip 
\noindent
{\bf II. HILL'S EQUATION WITH SMALL FLUCTUATIONS} 
\medskip 

In this section, we consider the case where the parameters of Hill's
equation are arbitrary, but the variations from cycle to cycle and/or
the amplitude of the stochastic process are small.

\medskip 
\noindent
{\bf Theorem 2.1:} Consider the CC Random Hill's equation
(\ref{ccrandom}) with the ordering of equation (\ref{ordering}). Let
$h$ and $g$ denote the matrix elements of the map (\ref{mapzero}), and
let $h_0$ and $g_0$ denote the matrix elements in the absence of
fluctuations ($\ell_k$ = 0 = $p_k$). Provided that the zeroeth order 
matrix elements $h_0$ and $g_0$ are nonvanishing, the matrix elements 
$h$ and $g$ are given by 
\be
h = h_0 - \ell_k X - p_k Y + {\cal O} (\epsilon^2) \qquad {\rm and} \qquad 
g = g_0 - \ell_k W - p_k Z + {\cal O} (\epsilon^2) \, , 
\label{melements} 
\ee
where $X,Y,W,Z$ are integral quantities that are constant from cycle 
to cycle:  
$$
X = {1 \over 2 g_0} \int_0^\pi 
\left[ \left( h_0^2 - 1 \right) y_{01}^2 - g_0^2 y_{02}^2 \right] dt \, , 
\qquad Y = {1 \over 2 g_0} \int_0^\pi \qhat(t) 
\left[ \left( h_0^2 - 1 \right) y_{01}^2 - g_0^2 y_{02}^2 \right] dt \, , 
$$
\be 
W = {1 \over 2 h_0} \int_0^\pi 
\left[ \left( h_0^2 + 1 \right) y_{01}^2 - g_0^2 y_{02}^2 \right] dt \, , 
\qquad Z = {1 \over 2 h_0} \int_0^\pi \qhat(t) 
\left[ \left( h_0^2 + 1 \right) y_{01}^2 - g_0^2 y_{02}^2 \right] dt \, .
\label{xyzwintegrals} 
\ee 

\medskip 
\noindent
{\it Proof:} 
By definition, the leading order part of the solution $y_0$ obeys the
Classical Hill's equation in the form of equation (\ref{classical}),
where $\lambda$ and $q$ are constant. To all orders (in $\epsilon$),
and for both principle solutions, the correction $\yp$ obeys a 
differential equation of the form 
\be
{d^2 \yp \over dt^2} + \left[ \lambda + \ell_k + (q + p_k) \qhat \right] 
\yp + \left[ \ell_k + p_k \, \qhat \right] y_0 = 0 \, ,
\ee
where $y_0$ is the zeroeth order part of the principal solution. 
To leading order in $\epsilon$, this equation reduces to the form 
\be
{d^2 \yp \over dt^2} + \left[ \lambda + q \qhat \right] 
\yp + \left[ \ell_k + p_k \, \qhat \right] y_0 + 
{\cal O} (\epsilon^2) = 0 \, . 
\label{work} 
\ee
For the remainder of this argument, we neglect terms that are 
${\cal O} (\epsilon^2)$, so that the results are correct (only) to
first order in $\epsilon$. After multiplying equation (\ref{work}) 
by $y_0$ and integrating over one cycle, the expression becomes
\be
\int_0^\pi {d^2 \yp \over dt^2} y_0 dt + 
\int_0^\pi \left[ \lambda + q \qhat \right] y_0 \yp dt + 
\int_0^\pi \left[ \ell_k + p_k \, \qhat \right] y_0^2 dt = 0 \, ,
\ee
where we ignore higher order terms. By integrating the first term 
by parts, applying the boundary conditions at $t$ = 0, and using 
the fact that $y_0$ satisfies the zeroeth order equation, we obtain 
\be
y_0(\pi) {\dot y}_k(\pi) - {\dot y}_0(\pi) \yp(\pi) + 
\ell_k \int_0^\pi y_0^2 dt + p_k \int_0^\pi \qhat y_0^2 dt = 0 \, . 
\label{inparts} 
\ee
This equation holds for both cases, i.e., where $y = y_0 + \yp$ is the
first or the second principal solution. Using the definitions of the 
matrix elements, we obtain two equations for two unknowns (note that 
we suppress the index $k$ as convenient to simplify the notation)
\be
h_0 g - g_0 h + K_1 = 0 \, , 
\ee
and 
\be
{h_0^2 - 1 \over g_0} h - h_0 {h^2 - 1 \over g} + K_2 = 0 \, , 
\ee
where the integrals $K_1$ and $K_2$ are defined by 
\be
K_1 \equiv \int_0^\pi \left[ \ell_k + p_k \, \qhat \right] y_{01}^2 dt 
\qquad {\rm and} \qquad 
K_2 \equiv \int_0^\pi \left[ \ell_k + p_k \, \qhat \right] y_{02}^2 dt \, . 
\ee
The matrix element $h$ is thus given by the quadratic equation 
\be
h^2 + h \left[ {h_0^2 - 1 \over g_0} K_1 - g_0 K_2 \right] 
+ K_1 K_2 - h_0^2 = 0 \, , 
\ee
which has solution 
\be
h = \pm \left[ h_0^2 - K_1 K_2 + {1 \over 4} 
\left( {h_0^2 - 1 \over g_0} K_1 - g_0 K_2 
\right)^2 \right]^{1/2} - 
{1 \over 2} \left(  {h_0^2 - 1 \over g_0} K_1 - g_0 K_2 \right) \, . 
\ee
We choose the $+$ solution so that $h \to h_0$ in the limit
$\ell_k,p_k\to0$.  This expression can be simplified by keeping only
the leading order terms in $\epsilon$, i.e.,
\be
h = h_0 - 
{1 \over 2} \left(  {h_0^2 - 1 \over g_0} K_1 - g_0 K_2 \right) 
+ {\cal O} (\epsilon^2) \, . 
\ee
A similar result can be obtained for $g$, i.e., 
\be
g = g_0 - {1 \over 2 h_0} \left[ (h_0^2 + 1) K_1 - g_0^2 K_2 \right]
+ {\cal O} (\epsilon^2) \, . 
\ee
Since we want to isolate the dependence of the results on the
parameters $\ell_k$ and $p_k$ that vary from cycle to cycle, the
matrix elements can be rewritten to take the forms given in equations
(\ref{melements}) and (\ref{xyzwintegrals}).  $\square$

Keep in mind that only the $(\ell_k,p_k)$ vary from cycle to cycle. 
As a result, the parameters $X,Y,W,Z$ are constant, i.e., the same for
all cycles.

Alternatively, we can define the moment integrals
\be
I_j = \int_0^\pi y_{0j}^2 dt \qquad {\rm and} \qquad 
J_j = \int_0^\pi \qhat(t) y_{0j}^2 dt \, , 
\label{moments} 
\ee
where $j$ = 1,2, and can write the parameters $X,Y,W,Z$ 
in the form 
$$
X = {1 \over 2 g_0} \left[\left(h_0^2 - 1\right) I_1 - g_0^2 I_2\right]\,, 
\qquad 
Y = {1 \over 2 g_0} \left[\left(h_0^2 - 1\right) J_1 - g_0^2 J_2\right]\,, 
$$
\be 
W = {1 \over 2 h_0} \left[\left(h_0^2 + 1\right) I_1 - g_0^2 I_2\right]\,, 
\qquad {\rm and} \qquad 
Z = {1 \over 2 h_0} \left[\left(h_0^2 + 1\right) J_1 - g_0^2 J_2\right]\,.
\label{xyzwalt} 
\ee 

\medskip
\noindent
{\bf Remark:} Note that this perturbation scheme fails in the limit
where the leading order principal solutions vanish, i.e., when either 
$h_0 \to 0$ or $g_0 \to 0$. 

The following results consider the case where $|h_0| < 1$ so that the
solutions are classically stable (i.e., stable in the absence of
fluctuations in the parameter values). For the case of nonzero
parameter fluctuations, however, the solutions grow exponentially.

\medskip 
\noindent
{\bf Theorem 2.2:} For a CC Random Hill's equation in the regime
where the leading order matrix element $|h_0|<1$, the solutions are
stable in the limit $\ell_k,p_k \to 0$. In general, the solutions are
unstable with growth rate 
\be
\gamma \approx {1 \over 2} (1 - h_0^2) \left\{ 
\left[ {h_0 \over 1 - h_0^2} X + {1 \over g_0} W \right]^2 
\langle \ell_k^2 \rangle 
+ \left[ {h_0 \over 1 - h_0^2} Y + {1\over g_0} Z \right]^2 
\langle p_k^2 \rangle \right\} + {\cal O} \left( \epsilon^3 \right) \, ,
\label{ccgrowth} 
\ee
where the angular brackets denote averages over the parameter
distributions, and where the integral quantities $X,Y,Z,W$ are defined
by equations (\ref{xyzwintegrals}).

\medskip 
\noindent
{\it Proof:} In the limit $\ell_k,p_k\to0$, the matrix element 
$h \to h_0$. Since $|h_0| < 1$, the solutions are stable [22]. 
For nonzero fluctuations, the transfer matrix (\ref{mapzero}) 
takes the form of an elliptical rotation matrix [4], which 
can be written in the form 
\be
{\mfont M}_k = {\mfont E} (\theta_k, L_k) = 
\left[ \matrix{\cos\theta_k & -L_k \sin\theta_k \cr 
(1/L_k)\sin\theta_k & \cos\theta_k} \right] \, , 
\label{elliptical} 
\ee
where $\cos\theta_k=h_k$ and $L_k=(\sin\theta_k)/g_k$. 
Theorem 4 of Ref. [4] shows that the growth rate for multiplication of
elliptical rotation matrices of the form (\ref{elliptical}), and 
hence the growth rate for the CC Random Hill's equation under 
consideration, can be written 
\be
\gamma = {1 \over 2} \langle \sin^2 \theta_k \rangle 
\langle \eta_k^2 \rangle + {\cal O} \left( \epsilon^3 \right) \, , 
\label{growold} 
\ee
where the perturbation $\eta_k$ is defined by 
$L_k = L_0 (1 + \eta_k)$.  

The length parameter $L_k$ of the elliptical rotation is given by 
\be
L_k = {(1 - h_k^2)^{1/2} \over g_k} \, . 
\label{lengthdef} 
\ee
Using the expressions for the matrix elements from equation
(\ref{melements}) in the definition (\ref{lengthdef}) of $L_k$, 
and keeping only the leading order terms in $\epsilon$, we find 
\be
L_k = {(1 - h_0^2)^{1/2} \over g_0} \left\{ 1 + {h_0 \over 1 - h_0^2} 
\left[ \ell_k X + p_k Y \right] + {1 \over g_0} 
\left[ \ell_k W + p_k Z \right] \right\} + 
{\cal O} \left( \epsilon^2 \right) \, , 
\ee
where $X,Y,Z,W$ are the integral quantities defined by equations
(\ref{xyzwintegrals}) or (\ref{xyzwalt}).  We can thus write the
length parameter $L_k$ in the form $L_k$ = $L_0 (1+\eta_k)$, where
\be
L_0 = {(1 - h_0^2)^{1/2} \over g_0} \qquad {\rm and} \qquad 
\eta_k = {h_0 \over 1 - h_0^2} \left[ \ell_k X + p_k Y \right] + 
{1 \over g_0} \left[ \ell_k W + p_k Z \right] \, . 
\ee
Since we have defined the parameters $\ell_k$ and $p_k$ (which 
vary from cycle to cycle) to have zero mean, it follows that 
$\langle \eta_k \rangle$ = 0. The quantity that appears in the 
growth rate from equation (\ref{growold}) is the expectation value
$\langle\sin^2\theta_k\rangle \langle\eta_k^2\rangle$. 
The perturbation $\eta_k$ = ${\cal O}(\epsilon)$, whereas to leading
order $\sin^2\theta_k$ = $1 - h_0^2 + {\cal O}(\epsilon)$. As a result, 
the growth rate can be written in the form 
\be
\gamma = {1\over2} (1 - h_0^2) 
\left\langle\left( \ell_k \left[ {h_0 \over 1 - h_0^2} X + 
{1 \over g_0} W \right] + p_k \left[ {h_0 \over 1 - h_0^2} Y + 
{1\over g_0} Z \right] \right)^2 \right\rangle + 
{\cal O} \left( \epsilon^3 \right) \, . 
\ee 
The random variables $\ell_k$ and $p_k$ are independently distributed,
so that averages over the distributions can be rewritten in the form 
$$
\left\langle\left( \ell_k \left[ {h_0 \over 1 - h_0^2} X + 
{1 \over g_0} W \right] + p_k \left[ {h_0 \over 1 - h_0^2} Y + 
{1\over g_0} Z \right] \right)^2 \right\rangle = 
\qquad \qquad \qquad \qquad \qquad \qquad 
$$
\be
\qquad \qquad \qquad \qquad \qquad \qquad 
\left[ {h_0 \over 1 - h_0^2} X + {1 \over g_0} W \right]^2
\langle \ell_k^2 \rangle + \left[ {h_0 \over 1 - h_0^2} Y + 
{1 \over g_0} Z \right]^2 \langle p_k^2 \rangle \, . 
\ee 
The growth rate $\gamma$ thus has the form given by equation 
(\ref{ccgrowth}).
$\square$ 

\medskip
\noindent
{\bf Theorem 2.3:} Consider a Stochastic Hill's equation of the form
(\ref{stochastone}) and a CC Random Hill's equation of the form
(\ref{ccrandom}), where the parameters have the forms $\af+\ell_k$ and
$q+p_k$.  Both the parameter variations ($\ell_k,p_k$) and the
stochastic process $\xi$ are ${\cal O}(\epsilon)$. To leading order in
$\epsilon$, the growth rates of the solutions of the two types of
differential equations are equivalent when the parameter variations 
($\ell_k,p_k$) are chosen according to 
\be
\ell_k = {J_2 \Xi_1 - J_1 \Xi_2 \over I_1 J_2 - I_2 J_1} 
\qquad {\rm and} \qquad 
p_k = {I_1 \Xi_2 - I_2 \Xi_1 \over I_1 J_2 - I_2 J_1} \, , 
\label{equivalence} 
\ee
where the integrals $I_j$ and $J_j$ are defined by equation
(\ref{moments}), and where the $\Xi_j$ are integrals of the 
stochastic process defined by 
\be
\Xi_1 \equiv \int_0^\pi y_{01} \xi dt 
\qquad {\rm and} \qquad 
\Xi_2 \equiv \int_0^\pi y_{02} \xi dt \, . 
\label{ximoments} 
\ee
The growth rates $\gamma$ are given by equation (\ref{ccgrowth}). 

\medskip
\noindent
{\it Proof:} The growth rates of both types of differential equations
can be written in terms of the growth rates of the transfer matrix
(under matrix multiplication). The growth rates will be the same if
the matrix elements are the same (see equation (\ref{mapzero})).

The Stochastic Hill's equation (\ref{stochastone}) has a solution of
the form given by equation (\ref{smallfunction}), where $y_0(t)$ is
the solution to equation (\ref{stochastone}) for the case where the
right hand side vanishes. We consider the stochastic process $\xi$ and
the correction $\yp$ to be ${\cal O}(\epsilon)$. To consistent order, 
$\yp(t)$ must obey the equation
\be
{d^2 \yp \over dt^2} + [ \af + q \qhat (t) ] \yp = \xi \, , 
\label{onesto} 
\ee
If we multiply the above equation by $y_0(t)$ and then integrate 
over one period, from $t=0$ to $t=\pi$, we obtain the result 
\be
y_0(\pi) {\dot y}_k(\pi) - {\dot y}_0(\pi) \yp(\pi) = 
\int_0^\pi y_0 \xi dt \, . 
\label{inpartsto} 
\ee
Now compare this result to that obtained from the CC Random Hill's 
equation where the parameters vary from cycle to cycle, 
\be
y_0(\pi) {\dot y}_k(\pi) - {\dot y}_0(\pi) \yp(\pi) + 
\ell_k \int_0^\pi y_0^2 dt + p_k \int_0^\pi y_0^2 \qhat dt 
\, = {\cal O} \left( \epsilon^2 \right) \, . 
\label{inpartran} 
\ee
The matrix elements of the transfer matrices will be the same if
equations (\ref{inpartsto}) and (\ref{inpartran}) are equal. But we
obtain equations of these forms for both principle solutions, so this
requirement results in two coupled equations
\be 
\ell_k I_1 + p_k J_1 = \Xi_1 \qquad {\rm and} \qquad 
\ell_k I_2 + p_k J_2 = \Xi_2 \, , 
\label{coupletwo} 
\ee
where where $I_j$ and $J_j$ are the moments defined in equation
(\ref{moments}) and the $\Xi_j$ are moments of the stochastic process
defined through equations (\ref{ximoments}).  

Equations (\ref{coupletwo}) provide two equations for two unknowns,
and can be solved to find the perturbative quantities $\ell_k$ and
$p_k$. We thus find the constraint of equation (\ref{equivalence}) as
claimed.  The moment integrals $I_j$ and $J_j$ are defined by the
solutions to the classical Hill's equation and are the same for all
cycles. On the other hand, the moments $\Xi_j$ of the stochastic
process will, in general, have different values for each realization
(over each cycle labeled by the index $k$). For a given cycle,
equations (\ref{equivalence}) thus define the values of the $\ell_k$
and $p_k$ that make the matrix elements for the CC Random Hill's
equation that same as those for the Stochastic Hill's equation. This
procedure thus results in distributions for the $\ell_k$ and $p_k$.
Finally, with the distributions of the (equivalent) $\ell_k$ and 
$p_k$ specified, the growth rate $\gamma$ is given by equation 
(\ref{ccgrowth}). $\square$

\medskip
\noindent
{\bf Corollary:} Consider a stochastic Hill's equation in the alternate form 
\be
{d^2 y \over dt^2} + [ \af + \left( q + \xi \right) 
\qhat (t) ] y = 0 \, . 
\label{altstochastic} 
\ee
The solutions to this differential equation will have the same growth
rate as that of the CC Random Hill's equation when the parameter
variations ($\ell_k,p_k$) are chosen according to the constraints of
equation (\ref{equivalence}), where the moments $\Xi_j$ have the
alternate form
\be
\Xi_1 = - \int_0^\pi \qhat y_{01}^2 \xi dt 
\qquad {\rm and} \qquad  
\Xi_2 = - \int_0^\pi \qhat y_{02}^2 \xi dt \, . 
\label{altximoments} 
\ee
{\it Proof:} The proof is analogous to that of the previous result. $\square$

\medskip
\noindent
{\bf Remark:} In the limit where the stochastic process $\xi$ has a
correlation time $\tau_c \to 0$, the integrals $\Xi_j$ in the numerators
of equations (\ref{equivalence}) vanish. As a result, the equivalent 
perturbations vanish, 
\be
\lim_{\tau_c \to 0} \ell_k = 0 
\qquad {\rm and} \qquad 
\lim_{\tau_c \to 0} p_k = 0 \, . 
\ee
In this limit, the growth rate of either differential equation will 
be the same as that of the corresponding Classical Hill's equation. 

\bigskip 
\noindent
{\bf III. HILL'S EQUATION WITH SMALL FORCING PARAMETERS}  
\medskip 

This section considers the CC Random Hill's equation (\ref{ccrandom})
in the limit where the natural oscillation frequency $\af$ is constant
and the forcing parameter $q_k$ = ${\cal O}(\epsilon)$ where
$\epsilon\ll{1}$. This application is thus a special case of the
general problem considered in the previous section.

We first find solutions for the matrix elements $h=y_1(\pi)$ and
$g={\dot y}_1(\pi)$. The solutions can be expanded in orders of 
$q_k = {\cal O}(\epsilon)$, where the zeroeth order solutions are
given by
\be
y_{01} (t) = \cos \lamroot t \, , \qquad {\rm and} \qquad 
y_{02} (t) = {1 \over \lamroot} \sin \lamroot t \, . 
\label{zero} 
\ee
The first order parts of the solutions obey the equation 
\be
{d^2 \yp \over dt^2} + \lambda \yp + q_k \qhat y_0 = 
{\cal O} (\epsilon^2) \,, 
\ee
where the parameter $q_k={\cal O}(\epsilon)$. 
Next we multiply by $y_0$ and integrate over the interval $[0,\pi]$.
After integrating by parts twice, the expression becomes
\be
\Bigl[ y_0 {\dot \yp} - {\dot y}_0 \yp \Bigr]_0^\pi + 
q_k \int_0^\pi \qhat y_0^2 dt = {\cal O} (\epsilon^2) \, . 
\label{solution} 
\ee
Note that this form holds for the perturbations for both the first and
second principal solutions. Since the zeroeth order solutions (given
by equation (\ref{zero})) satisfy the boundary conditions, the
perturbations $\yp$ and their derivatives ${\dot \yp}$ must vanish at
$t$ = 0.  The solution thus becomes
\be 
y_{0j} (\pi) {\dot \yp}_j (\pi) - 
{\dot y}_{0j} (\pi) \yp_j (\pi) + q_k J_j = 0 \, , 
\label{reduce} 
\ee
where the index $j$ = 1,2 determines the principal solution and the
moment integrals $J_j$ are defined by equation (\ref{moments}). The
matrix elements $h$ and $g$ are given by
\be
h = y_1(\pi) = \cos\varphi + \yp_1 (\pi) \qquad {\rm and} \qquad  
g = {\dot y}_1(\pi) = -\lamroot \sin\varphi + {\dot \yp}_1 (\pi) \,, 
\ee
where we have defined $\varphi\equiv\lamroot\pi$ to simplify the
zeroeth order parts. Using these forms in the two equations
(\ref{reduce}), we find two expressions for the two unknown matrix
elements $h$ and $g$: 
\be 
g \cos \varphi + h \lamroot \sin \varphi + q_k J_1 = 
{\cal O} (\epsilon^2) \, , 
\ee
and 
\be
h {\sin \varphi \over \lamroot} - {h^2 - 1 \over g} \cos \varphi 
+ q_k J_2 = {\cal O} (\epsilon^2) \, . 
\ee
The matrix elements are thus given by 
\be
h = \cos \varphi - {q_k \over 2 \lamroot} \sin \varphi + 
{\cal O} (\epsilon^2) \, , 
\label{hestimate} 
\ee 
and 
\be
g = - \lamroot \sin \varphi + {q_k \over \cos \varphi} 
\left( {1 \over 2} \sin^2 \varphi - J_1 \right) + 
{\cal O} (\epsilon^2) \, . 
\label{gestimate} 
\ee
Note that the expression for $h$ is the same as the full solution for
$h$ in the limiting case where the barrier shape $\qhat$ is a delta
function [3]. In this same limit, the integral $J_1=\cos^2(\varphi/2)$,
and the expression for $g$ also gives the exact result. Notice also 
that the results of Theorem 2.1 reduce to equations (\ref{hestimate}) and 
(\ref{gestimate}) in the limit $\ell_k \to 0$ and $q + p_k \to q_k$ 
(where equations [\ref{zero}] specify the zeroeth order solutions). 

Note that when $\cos \varphi \to 0$, the expression for $g$ must be
evaluated carefully. Both the numerator and the denominator vanish, 
and the limit must be evaluated. These values occur when $\lambda = 
(k + 1/2)^2$, where $k \in {\cal Z}$.

\begin{figure} 
\centering
\includegraphics[scale=0.75]{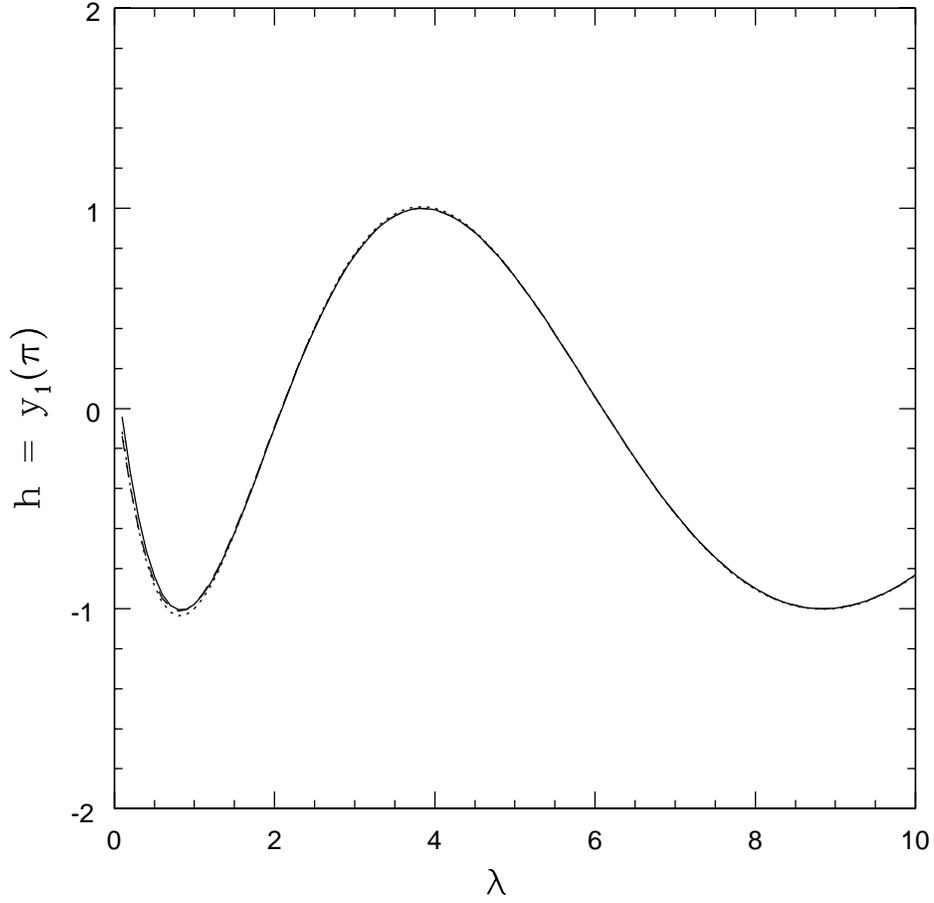} 
\caption{Matrix element $h$ as a function of natural oscillation 
frequency $\af$ for Hill's equation with small forcing parameter $q_k$
= 1/2. Solid curve shows numerically determined result. Dotted curve
shows the result for the approximation of equation (\ref{hestimate}),
which is correct to leading order in $q_k$. Dashed curve shows the
result for the approximation of equation (\ref{hmixed}), which
includes some higher order terms (see text). }
\label{fig:qhalf} 
\end{figure}

Figure \ref{fig:qhalf} illustrates the fidelity of the first order
approximation scheme.  The figure shows the matrix element $h$ as a
function of $\af$ using the exact (numerically determined) form and
the first order expression from equation (\ref{hestimate}). For this
illustration, the parameter $q_k$ has constant value $q_k$ = 1/2. Note
that the first order expression provides a good estimate. Smaller
$q_k$ values will result in an even better approximation.

For completeness, we note the following: Although the approximation
(from equation (\ref{solution})) ignores terms that are second order
in $q_k$, after we make that approximation we can find the corresponding
solution keeping all terms, i.e., we don't have to get rid of the
remaining second order terms. We thus find the generalized solutions 
\be
h = - {q_k \over 2 \sqrt{\lambda} } \sin \varphi \pm 
\left[ \cos^2 \varphi + {q_k^2 \over 2 \lambda} \sin^2 \varphi 
- q_k^2 J_1 J_2 \right]^{1/2} \, , 
\label{hmixed} 
\ee
and 
\be
g = {1 \over \cos\varphi} \left\{ - q_k J_1 + 
{q_k \over 2} \sin^2 \varphi \mp \sqrt{\lambda} \sin \varphi 
\left[ \cos^2 \varphi + {q_k^2 \over 2 \lambda} \sin^2 \varphi 
- q_k^2 J_1 J_2 \right]^{1/2} \right\} \, , 
\label{gmixed} 
\ee 
where the signs are chosen so that the expressions reduce to 
those of equations (\ref{hestimate}) and (\ref{gestimate}) 
when the $q_k^2$ terms are ignored. 

We now consider the growth rates.  The transformation matrix of
equation (\ref{mapzero}) can be written in the form of an elliptical
rotation matrix, as in equation (\ref{elliptical}). The length
parameter of the rotation can be written in the form $L_k$ = 
$L_0(1+\eta_k)$, where the $\eta_k$ vary from cycle to cycle. For
the case of symmetric variations, where $\langle\eta_k\rangle=0$, 
the growth rate takes the form
\be
\gamma = \log \left[ 1 + {1 \over 2} \langle \eta_k^2 \rangle 
\langle \sin \theta_k^2 \rangle \right] \, , 
\label{gammaold} 
\ee 
where $\theta_k$ is the angle of the elliptical rotation matrix
(see [4] and Theorem 2.2). 

Using the matrix elements found above, and dropping the subscripts
on the $h$ and $g$, we find 
\be
\cos \theta_k = h = \cos \varphi - {q_k \over 2 \lamroot} \sin \varphi 
+ {\cal O}(\epsilon^2) \, , 
\ee
so that 
\be
\theta_k = \varphi + {q_k \over 2 \lamroot} + {\cal O}(\epsilon^2) \, , 
\ee
and hence 
\be
\sin \theta_k = (1 - h^2)^{1/2} = \sin \varphi +  
{q_k \over 2 \lamroot} \cos \varphi + {\cal O}(\epsilon^2) \, . 
\ee
The length parameter $L_k$ of the elliptical rotation matrix 
can then be written in the form 
\be
L_k = {\sin \theta_k \over g} = - {1 \over \lamroot} 
\left[ 1 + {1 - 2 J_1 \over \cos\varphi \sin\varphi} 
{q_k \over 2 \lamroot} \right] \equiv - {1 \over \lamroot} 
\left[ 1 + \eta_k \right] + {\cal O}(\epsilon^2) \, ,
\label{lform} 
\ee
where the final equality in equation (\ref{lform}) defines the
perturbation $\eta_k$.  Combining with the above results we find 
the following Theorem: 

\medskip 
\noindent
{\bf Theorem 3.1:} The growth rate for the CC Random Hill's equation 
in the limit of constant $\af$ and small $q=q_k={\cal O}(\epsilon)$
is given by 
\be
\gamma = \log \left[ 1 + {\langle q_k^2 \rangle \over 8 \lambda} 
\left( {2 J_1 - 1 \over \cos\varphi} \right)^2  \right] \, 
\approx {\langle q_k^2 \rangle \over 8 \lambda} 
\left( {2 J_1 - 1 \over \cos\varphi} \right)^2 
+ {\cal O}(\epsilon^2) \, . 
\label{growsmallq} 
\ee
Note that this form is valid for the case of symmetric variations 
of the length parameter, i.e., where the distribution of the $\eta_k$  
is symmetric w.r.t. zero. 

Notice also that the expression for the growth rate is not defined 
when $\cos \varphi \to 0$. We can define the quantity $J$ such that 
\be
J = {2 J_1 - 1 \over \cos\varphi} = {1 \over \cos (\lamroot \pi) } 
\int_0^\pi \cos (2 \lamroot t) \qhat dt \, . 
\ee 
When $\cos \varphi \to 0$, $\lamroot \to (2k + 1)/2$, and both the 
numerator and denominator of the above expression vanish. To evaluate
the quantity $J$, we can use L'Hopital's rule, which implies that 
\be
J = \pm {2 \over \pi} \int_0^\pi t \sin [(2k+1) t] \qhat dt \, , 
\ee
where the $+(-)$ sign arises for $k$ even(odd). 

\begin{figure} 
\centering
\includegraphics[scale=0.75]{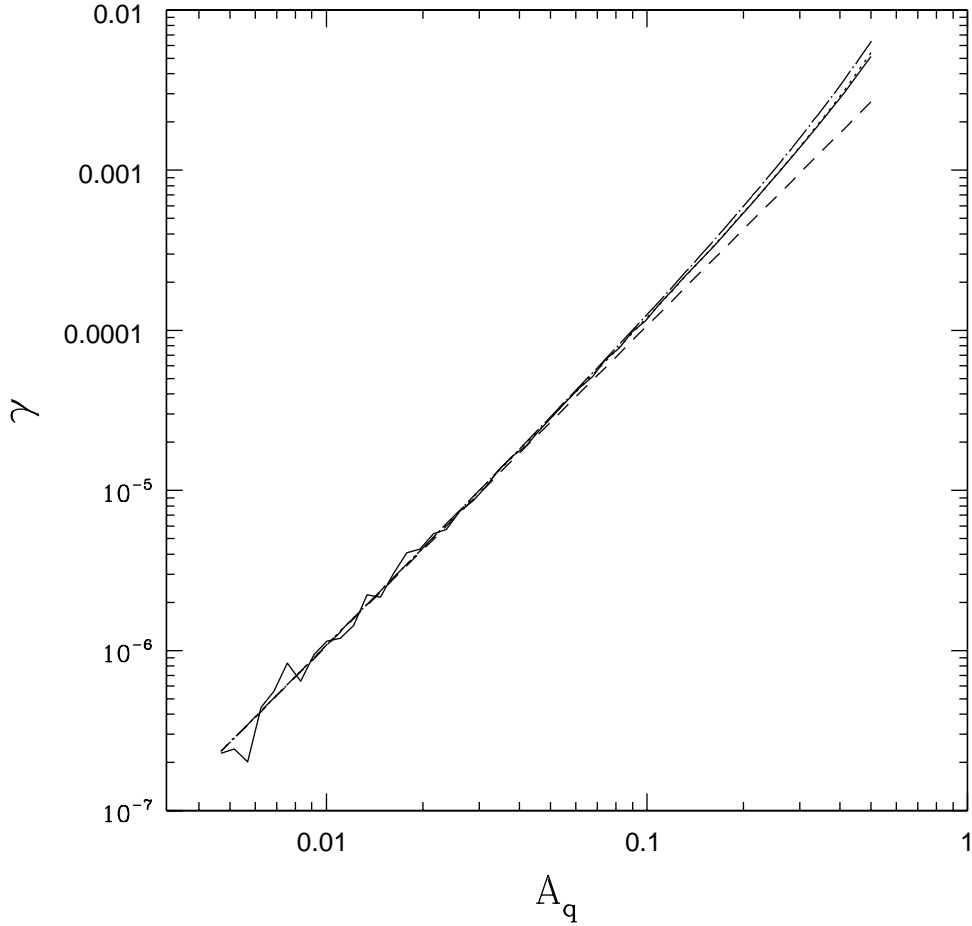} 
\caption{Growth rates for random Hill's equation in the limit of 
small random forcing parameter $q_k$, plotted as a function of the
amplitude $A_q$ of the fluctuations (for $\af$ = 1/2).  The solid
curve shows the result from direct matrix multiplication; dotted curve
shows the result from equation (\ref{gammaold}), where the $\eta_k$
are sampled numerically; the dashed curve shows the result from
equation (\ref{growsmallq}); the dot-dashed curve shows the growth
rate calculated by including higher order terms (in $q_k$). }
\label{fig:growamp} 
\end{figure}

Figure \ref{fig:growamp} illustrates how well this approximation
scheme works.  For the sake of definiteness, the $q_k$ are allowed to
vary over a uniform distribution with amplitude $A_q$.  We expect the
above results for the growth rate to be exact in the limit of small
fluctuations, i.e., where the fluctuation amplitude $A_q \to 0$.  The
figure shows the growth rates as a function of the amplitude $A_q$ for
constant $\af$ = 1/2.  Here the matrix elements are calculated using
the first order expressions from equations (\ref{hestimate}) and
(\ref{gestimate}).  The solid curve shows an estimate of the growth
rate determined directly from matrix multiplication; note that the
growth rate converges very slowly at small amplitudes, so the plot
contains errors due to incomplete sampling. The dotted curve shows the
growth rate calculated from equation (\ref{gammaold}), where the
variations $\langle \eta_k^2 \rangle$ are found by numerical
sampling. The dashed curve shows the growth rate from equation
(\ref{growsmallq}), which represents the main result of this
section. Finally, even though the expressions derived here are correct
only to first order in $q_k$, we can include the second order terms 
(in $q_k$) for purposes of determining the $\eta_k$ and hence the
growth rate; for this case, the result is shown as the dot-dashed
curve. All four of the curves are coincident for sufficiently small
amplitudes $A_q$.  The analytic expression of equation
(\ref{growsmallq}) represents the crudest approximation (dashed
curve), and converges the slowest.

\medskip 
\noindent 
{\bf Corollary:} Consider a CC Random Hill's equation in the limit of
small (but finite) forcing strength $|q_k| \ll 1$, with constant
oscillation parameter $\af$, and where $\qhat(t)$ is a function.  In
the limit $\af \to \infty$, the growth rate $\gamma \to 0$.

\medskip 
\noindent
{\it Proof:} In this limit, the growth rate is given by equation
(\ref{growsmallq}). For nonzero fluctuations $\langle q_k^2 \rangle > 0$, 
the growth rate vanishes if and only if the integral $J_1$ = 1/2, i.e., 
\be
J_1 = \int_0^\pi dt \, \qhat(t) \cos^2 (\sqrt{\af} t) = 1/2 \, .
\ee 
Using trigonometric identities, this expression can be 
written in the form 
\be
\int_0^\pi dt \, \qhat(t) \cos (2 \sqrt{\af} t) = 0  \, .  
\label{intozero} 
\ee
In the limit $\af \to \infty$, this integral vanishes, and hence 
$\gamma \to 0$. $\square$

\medskip
\noindent 
{\bf Remark:} The above argument works provided that $\qhat(t)$ is a
function.  For the limiting case where $\qhat$ is a delta-function
(and hence a distribution), the integral of equation (\ref{intozero})
does not necessarily vanish; nonetheless, the growth rate $\gamma
\propto 1/\af$ in this limit [3], so that $\gamma \to 0$ as 
$\af \to \infty$. 

For example, let $\qhat = (2/\pi) \sin^2 t$. Then the quantity 
$2 J_1 - 1$ becomes 
\be
2J_1 - 1 = - {\sin 2 \varphi \over 2 \pi \sqrt{\af} (\af - 1) } \, , 
\ee 
which vanishes as $\af \to \infty$. 

\begin{figure} 
\centering
\includegraphics[scale=0.75]{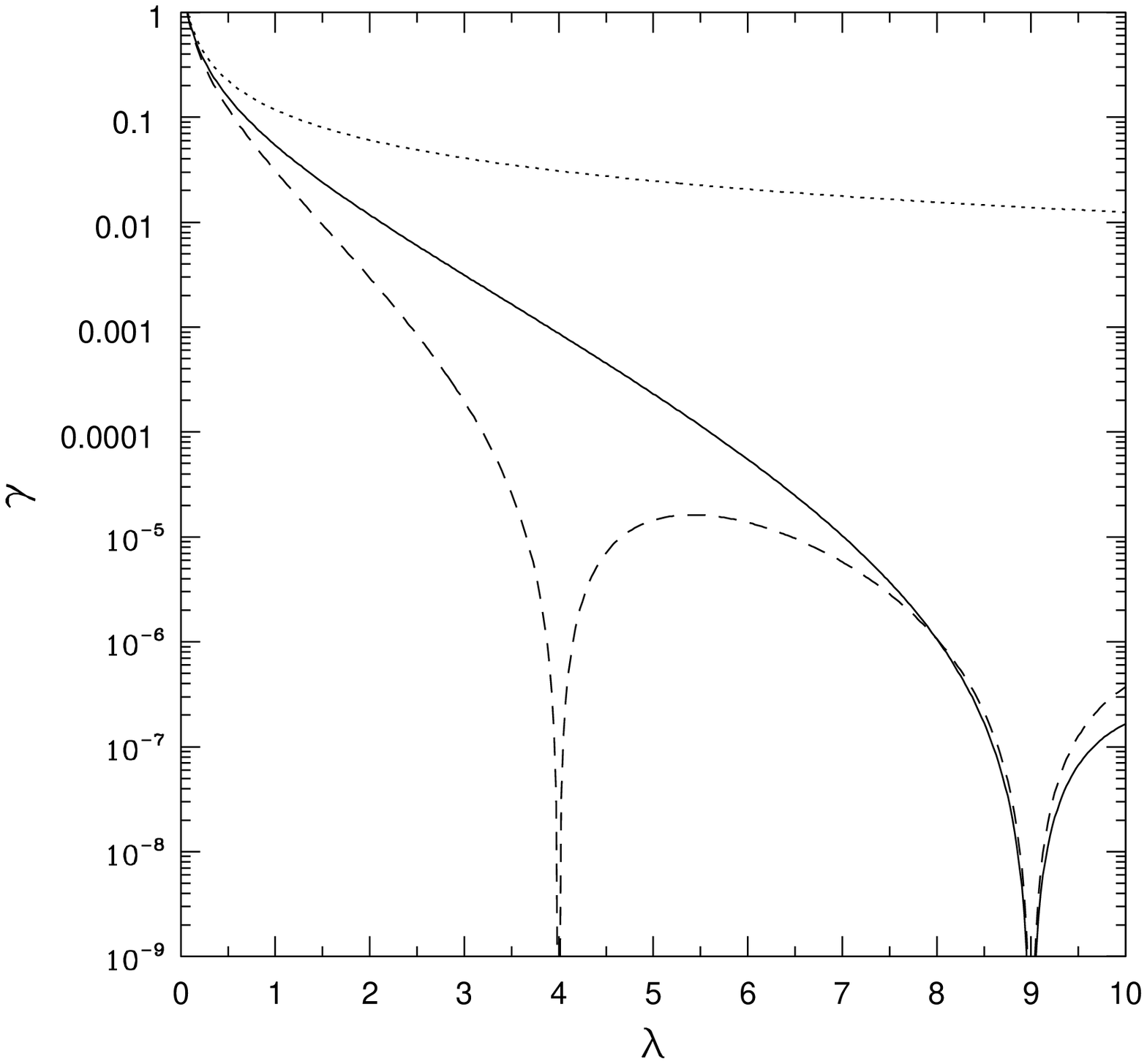} 
\caption{Growth rates for random Hill's equation as a function of 
$\af$, where the fluctuation amplitude $\langle q_k^2 \rangle$
= 1. The three curves show the result for $\qhat=\delta([t]-\pi/2)$ 
(dotted), $\qhat = (3/8\pi) \sin^4 t$ (solid), and for
$\qhat=(2/\pi)\sin^2 t$ (dashed). }
\label{fig:gamvlam} 
\end{figure}

The general trend of decreasing growth rate $\gamma$ with increasing
$\af$ is shown in Figure \ref{fig:gamvlam} for three choices for the
barrier function. The various curves show the growth rates for
$\qhat(t)=(8/3\pi)\sin^4 t$ (solid), $\qhat(t)=(2/\pi)\sin^2 t$
(dashed), and $\qhat(t) = \delta ([t] - \pi/2)$ (dotted), where we
have included the normalization constants, and where $[t]$ denotes
that the variable $t$ is to be evaluated mod-$\pi$. The amplitude 
of the $q_k$ fluctuations are the same for all three cases. 

\bigskip 
\noindent{\bf IV. CONCLUSION} 
\medskip

This paper has generalized the Classical Hill's equation
(\ref{classical}) to include random elements, where this treatment
focuses on the case of small fluctuations with amplitude 
${\cal O}(\epsilon)$. This generalization can take two distinct, 
but related, forms: The parameters of the CC Random Hill's equation
(\ref{ccrandom}) vary from cycle to cycle, whereas the Stochastic
Hill's equation (\ref{stochastone}) or (\ref{stochasttwo}) includes 
a stochastic process.

Theorem 2.3 shows that the growth rates of these two types of
(generalized) Hill's equations are the same when (for a given
stochastic process $\xi$) the parameter variations of the CC Random
Hill's equation are chosen according to equation (\ref{equivalence}).
Theorem 2.2 provides an approximation for the growth rate, correct to
${\cal O}(\epsilon^2)$, for the CC Random Hill's equation (and hence
for the equivalent Stochastic Hill's equation). These growth rates
$\gamma\sim\langle{p_k^2}\rangle$ are nonzero except for isolated
cases (see Figure \ref{fig:gamvlam}).  As a result, for the CC Random
Hill's equation, the (generalized) plane of parameters has no regions
of stability. This result is not due to the fluctuations moving the
matrix elements into the unstable regions of parameter space. One
obtains nonzero growth rates even when all of the matrix elements are
classically stable. In contrast, the Stochastic Hill's equation can be
stable, where stability requires that the moments $\Xi_j$ of the
stochastic process vanish (see equations [\ref{ximoments}] and
[\ref{altximoments}]). Stability occurs in the limit where the
correlation time $\tau_c \to 0$.

The results of this paper can be used in a wide variety of
applications.  As one example, orbit instabilities in extended mass
distributions, such as dark matter halos, lead to a CC Random Hill's
equation, as shown in Appendix A (see also Refs. [5,12]).  As another
example, the reheating problem at the end of inflation naturally leads
to a Stochastic Hill's equation, as shown in Appendix B (see also
Refs. [17,18,27]), and the growth rates can be calculated using
Theorems 2.2 and 2.3. 

%\newpage 
\bigskip 
\bigskip 
\noindent
{\bf ACKNOWLEDGMENTS}
\medskip 

We thank Jake Ketchum and Jeff Lagarias for useful conversations.  
FCA and AMB are jointly supported by NSF Grant DMS-0806756 from the
Division of Applied Mathematics, and by the Michigan Center for
Theoretical Physics. AMB is also supported by NSF grants DMS-0907949
and DMS-1207693; FCA is also supported by NASA grant NNX11AK87G.

\bigskip 
\bigskip 
\renewcommand{\theequation}{A\arabic{equation}}
\setcounter{equation}{0}  % reset counter 

\noindent{\bf APPENDIX A: } 

\noindent{\bf RANDOM HILL'S EQUATION FROM ASTROPHYSICAL ORBITS} 
\bigskip 

Orbits in extended mass distributions are subject to an instability
[5] that can be described by a CC Random Hill's equation
(\ref{ccrandom}).  As one example, the density profile $\rho(\varpi)$
of a dark matter halo has the form
\be 
\rho (\varpi) = \, \rho_0 \, {F(\varpi) \over \varpi} \, , 
\label{rhogeneral} 
\ee 
where $\rho_0$ is a density scale and the variable $\varpi$ is written
in terms of Cartesian $(x,y,z)$ coordinates through the relation
\be
\varpi^2 = {x^2 \over a^2} + {y^2 \over b^2} + {z^2 \over c^2} \, , 
\label{mdef} 
\ee 
where (without loss of generality) $a > b > c > 0$.  The density field
is thus constant on ellipsoids. The function $F(\varpi)$ approaches
unity in the limit $\varpi \to 0$ so that the density profile
approaches the form $\rho \sim 1/\varpi$. For this regime, one can
find analytic forms for both the potential and the force terms [5].

When an orbiting body is initially confined to any of the three
principal planes, the motion can be unstable to perturbations in the
perpendicular direction. Consider an orbit initially confined to the
$x-z$ plane, with a small perturbation in the perpendicular $\hat y$
direction.  In the limit $|y| \ll 1$, the equation of motion for the
$y$-coordinate takes the form
\be 
{d^2 y \over dt^2} + \Omega_y^2 y = 0 \qquad {\rm where} \qquad 
\Omega_y^2 = { 4/b \over \sqrt{c^2 x^2 + a^2 z^2} + b \sqrt{x^2 + z^2} } \ . 
\label{omegay} 
\ee 
In this context, the time evolution of the coordinates $(x,z)$ is
determined by the original orbit and the motion is nearly periodic.
As a result, the $[x(t),z(t)]$ dependence of the parameter
$\Omega_y^2$ provides a (nearly) periodic forcing term. The orbit has
outer turning points which define a minimum value for $\Omega_y^2$,
which in turn defines the natural oscillation frequency $\af_k$. The
function $\Omega_y^2$ defined above can be written in the form
\be
\Omega_y^2 = \af_k + Q_k (t) \, , 
\label{expandomega} 
\ee
where the index $k$ counts the number of orbit crossings, and the
chaotic orbit in the original plane leads to different $\af_k$ and
$Q_k(t)$ for each crossing. The shape of the functions $Q_k(t)$ are
nearly the same, so that one can write $Q_k (t) = q_k \qhat(t)$, where
the forcing strength parameters $q_k$ vary from cycle to cycle. These
forcing strengths $q_k$ are determined by the inner turning points of
the orbit (weighted by the axis parameters $[a,b,c]$).  Given the
expansion of equation (\ref{expandomega}), the equation of motion
(\ref{omegay}) for the perpendicular coordinate becomes a CC Random
Hill's equation, with the form of equation (\ref{ccrandom}). 

\bigskip 
\bigskip 
\renewcommand{\theequation}{B\arabic{equation}}
\setcounter{equation}{0}  % reset counter 

\noindent{\bf APPENDIX B: } 

\noindent{\bf STOCHASTIC HILL'S EQUATION FROM REHEATING IN INFLATION} 
\bigskip 

In the inflationary universe paradigm [14], the accelerated expansion
of the universe is driven by the vacuum energy associated with a
scalar field $\varphi$ (or fields). During the phase of accelerated
expansion, the energy density of the universe itself decreases
exponentially and the cosmos becomes increasingly empty. This epoch is
thought to take place when the universe is extremely young, with
typical time scales of $\sim 10^{-36}$ sec [19].  In order for the
inflationary epoch to solve the cosmological issues it was designed to
alleviate, the end of inflation must include a mechanism to refill the
universe with energy [19].  This process is called reheating or
preheating.

During the reheating epoch, the equation of motion for the inflation
field displays oscillatory behavior about the minimum of its
potential. In order for the universe to become filled with energy
(reheat), the inflation field $\varphi$ must couple to matter or
radiation fields.  One simple type of interaction arises from a
coupling term in the Lagrangian of the form
\be
{\cal L}_{\rm int} = g \varphi \chi^2 \, , 
\ee
where $\chi$ is another scalar field that represents matter (or
radiation) and where the coupling constant $g$ sets the interaction
strength. The field $\chi$ can be expanded in terms of its Fourier
modes $\chi_\ell$ since these quantities evolve independently.  The
resulting equation of motion for the matter field modes $\chi_\ell$ 
then takes the form 
\be
{d^2 \chi_\ell \over dt^2} + 
\left[ \Omega_\ell^2 + Q_\ell(t) + \xi \right] \chi_\ell = 0 \, , 
\label{reheat} 
\ee
where $Q_\ell(t)$ is a periodic function (given by oscillatory
behavior of the inflation field) and $\xi$ is a noise term that can be
described by a stochastic process. Keep in mind that the index $\ell$
refers to the Fourier mode.  In the absence of fluctuations, the modes
$\chi_\ell$ of the matter fields obey a Classical Hill's equation
(\ref{classical}), which can be subject to parametric instability
[17,18]. However, the noise perturbations $\xi$ [17,18,27] convert the
reheating equation (\ref{reheat}) into a stochastic Hill's equation.
As shown by Theorem 2.3, the growth rates for this stochastic Hill's
equation are the same as an equivalent CC Random Hill's equation,
where the conditions for equivalence are given by equation
(\ref{ximoments}); the growth rates can thus be calculated according
to Theorem 2.2.

\bigskip 
\bigskip 
\newpage 
\noindent 
{\bf REFERENCES} 
\bigskip 

\medskip\noindent
[1] Abramowitz, M., and Stegun, I. A., 
{\it Handbook of Mathematical Functions} (Dover, New York, 1970). 

\medskip\noindent
[2] Adams, F. C.,  and Bloch, A. M., 
``Hill's Equation with random forcing terms,''
SIAM J. Appl. Math. {\bf 68}, pp. 947 -- 980 (2008).  

\medskip\noindent
[A3] Adams, F. C., and Bloch, A. M., ``Hill's Equation with 
random forcing terms: The limit of delta function barriers,'' 
J. Math. Phys. {\bf 50}, 073501 (2009).

\medskip\noindent
[4] Adams, F. C., and Bloch, A. M., ``Hill's Equation with random
forcing parameters: General treatment including marginally stable
cases,'' J. Stat. Phys. {\bf 139} pp. 139 -- 158 (2010).

\medskip\noindent
[5] Adams, F. C., Bloch, A. M., Butler, S. C., Druce, J. M., and
Ketchum, J. A., ``Orbits and instabilities in a triaxial cusp potential,'' 
Astrophys. J. {\bf 670}, pp. 1027 -- 1047 (2007). 

\medskip\noindent
[6] Anderson, P. W., ``Absence of diffusion in certain random lattices,''
Physical Review {\bf 109}, pp. 1492 -- 1505 (1958). 

\medskip\noindent
[7] Binney, J., ``Resonant excitation of motion perpendicular to galactic 
planes,'' Mon. Not. Royal Astron. Soc. {\bf 196}, pp. 455 -- 467 (1981). 

\medskip\noindent
[8] Binney, J. and Tremaine, S., {\it Galactic Dynamics}, 
(Princeton Univ. Press, Princeton, 1987).

\medskip\noindent
[9] Cohen, J. E., and Newman, C. M., ``The stability of large random
matrices and their products,'' Annals of Prob. {\bf 12}, pp. 283 -- 310 (1984). 

\medskip\noindent
[10] Carmona, R., and Lacroix, J., {\it Spectral Theory of} 
{\it Random Schr{\"o}dinger Operators}, (Birkhauser, Boston, 1990). 

\medskip\noindent
[11] Chandrasekhar, S., ``Stochastic Problems in Physics and Astronomy,''
Rev. Mod. Phys. {\bf 15} pp. 1 -- 89 (1947).

\medskip\noindent
[12] Furstenberg, H., ``Noncommuting random products,'' Trans. Amer. 
Math. Soc. {\bf 108}, pp. 377 -- 428 (1963). 

\medskip\noindent
[13] Furstenberg, H., and Kesten, H., ``Products of random matrices,'' 
Ann. Math. Stat. {\bf 31}, pp. 457 -- 469 (1960). 

\medskip\noindent
[14] Guth, A. H., ``Inflationary Universe: A possible solution to the
horizon and flatness problems,'' Phys. Rev. D {\bf 23}, pp. 347 -- 356 (1981).

\medskip\noindent
[15] Hill, G. W., ``On the part of the motion of the lunar perigee
which is a function of the mean motions of the Sun and Moon,''
Acta. Math. {\bf 8}, pp. 1 -- 36 (1886). 

\medskip\noindent 
[16] Keller, J. B., ``Ponytail Motion,'' SIAM J. Appl. Math. {\bf 70}, 
pp. 2667 -- 2672 (2010).  

\medskip\noindent
[17] Kofman, L., Linde, A., and Starobinsky, A. A., 
``Reheating after Inflation,'' Phys. Rev. Lett. {\bf 73}, 
pp. 3195 -- 3198 (1994).

\medskip\noindent
[18] Kofman, L., Linde, A., and Starobinsky, A. A., 
``Towards the theory of reheating after Inflation,'' 
Phys. Rev. D {\bf 56}, pp. 3258 -- 3295 (1997). 

\medskip\noindent
[19] Kolb, E. W., and Turner, M. S., {\it The Early Universe}, 
(Addison-Wesley, Reading MA, 1990). 

\medskip\noindent
[20] Lima, R., and Rahibe, M.,
``Exact Lyapunov exponent for infinite products of random matrices,'' 
J. Phys. A. Math. Gen. {\bf 27}, pp. 3427 -- 3437 (1994). 

\medskip\noindent 
[21] Lubow, S. H. ``Tidally driven inclination instability in
Keplerian disks,'' Astrophys. J. {\bf 398}, pp. 525 -- 530 (1992).

\medskip\noindent
[22] Magnus, W., and Winkler, S., {\it Hill's Equation}, (Wiley, New York, 1966). 

\medskip\noindent
[23] Pastur, L., and Figotin, A., 
{\it Spectra of Random and Almost-Periodic Operators}, a Series of
Comprehensive Studies in Mathematics, (Springer-Verlag, Berlin, 1991).

\medskip\noindent
[24] Pincus, S. ``Strong laws of large numbers for products of random matrices,'' 
Trans. Amer. Math. Soc. {\bf 287}, pp. 65 -- 89 (1985). 

\medskip\noindent
[25] Poulin, F. J., and Flierl, G. R., ``The stochastic Mathieu's
equation,'' Proc. Royal Soc. A {\bf 464}, pp. 1885 -- 1904 (2008).

\medskip\noindent
[26] Van Kampen, N. G., {\it Stochastic Processes in Physics}
{\it and Chemistry}, (North Holland, Amsterdam, 2001). 

\medskip\noindent
[27] Zanchin, V., Maia, A., Craig, W., and Brandenberger, R., 
``Reheating in the presence of noise,'' Phys. Rev. D. {\bf 57},  
pp. 4651 -- 4662 (1998). 

\medskip\noindent
[28] Zounes, R. S., and Rand, R. H., ``Transition curves for the 
quasi-periodic Mathieu equation,'' SIAM J. Appl. Math. {\bf 58},  
pp. 1094 -- 1115 (1998). 

\end{document}